\documentstyle[epsf,prl,aps,floats]{revtex}
\begin{document}
\title{Spinning jets}

\author{J. Eggers$^1$ and M. P. Brenner$^2$}
\address{
$^1$Universit\"{a}t Gesamthochschule Essen, Fachbereich Physik, 
45117 Essen, Germany \\
$^2$Department of Mathematics, MIT, Cambridge, MA 02139 
   }

\maketitle
\begin{abstract}
A fluid jet with a finite angular velocity is subject to centripetal
forces in addition to surface tension forces. At fixed angular
momentum, centripetal forces become large when the radius of
the jet goes to zero. We study the possible importance of this 
observation for the pinching of a jet within a slender jet model. 
A linear stability analysis shows the model to break down at 
low viscosities. Numerical simulations indicate that angular momentum
is expelled from the pinch region so fast that it 
becomes asymptotically irrelevant in the limit of the neck radius 
going to zero.
\end{abstract}

\section{Introduction}
A fluid jet emanating from a nozzle will become unstable and
break up due to surface tension. Some 30 years ago, a series of 
papers \cite{H60,GK61,P67,RJ70} investigated the modifications 
Rayleigh's classical analysis would undergo if the jet 
performed a solid-body rotation. Such a rotation is easily 
imparted by spinning the nozzle at the appropriate frequency. 
The somewhat surprising result of the linear analysis is that 
the rotation always {\it destabilizes} the jet, a wavenumber 
$k$ being unstable if
\begin{equation}
0 < kr_0 < (1 + L^{-1})^{1/2} ,
\label{stab}
\end{equation}
with 
\begin{equation}
L = \gamma/(\rho \Omega^2 r_0^3) .
\label{l}
\end{equation}
Here $\gamma$ is the surface tension, $\rho$ the density,
$\Omega$ the angular frequency, and $r_0$ the unperturbed jet 
radius. Note that $\Omega$ appears in the denominator, so 
no rotation corresponds to $L \rightarrow\infty$, for 
which the stability boundary 
$0 < k r_0 < 1$ found by Plateau is recovered. The theoretical
growth rates were found to be in reasonable agreement with
experiment \cite{RJ70} and growth of disturbances for $kr_0$
larger than 1 was confirmed. 

Recently it was pointed out by Nagel \cite{N98} that rotation 
might have an even more dramatic effect for the highly nonlinear 
motion near the point where the neck radius goes to zero.
Assume for the sake of the argument that a cylinder of fluid
of length $w$ pinches uniformly, i.e. it retains its cylindrical 
shape. Then the total angular momentum is 
\begin{equation}
M = \frac{\pi}{2} \rho\Omega r_0^4 w,
\label{m}
\end{equation}
and the volume $V = \pi r_0^2 w$ is constrained to remain
constant as $r_0$ goes to zero. The total interface pressure
corresponding to the outward centripetal force is found to 
be $p_c = \rho r_0^2\Omega^2/2$, and thus 
\begin{equation}
p_c = 2M^2/(V^2\rho r_0^2).
\label{pc}
\end{equation}
As $r_0$ goes to zero, this outward pressure will {\it dominate}
the surface tension pressure $\gamma/r_0$, raising the possibility 
that rotation is a singular perturbation for pinching: an arbitrarily
small amount of angular momentum, caused by a symmetry breaking, 
could modify the breaking of a jet. 

However, a jet does not pinch uniformly, but rather in a 
highly localized fashion \cite{E97}. If the above argument is 
applied to pinching, it must correspond to a rapidly spinning
thread of fluid surrounded by almost stationary fluid. 
Frictional forces represented by the viscosity of the fluid will
lead to a diffusive transport of angular momentum out of the 
pinch region, thus reducing its effect. Determining which
effect dominates requires a fully nonlinear calculation, 
including effects of surface tension, viscosity, inertia, 
and centripetal forces. In the spirit of earlier work for 
$\Omega = 0$ \cite{ED94,BFL} we will derive a one dimensional 
model, which only takes into account the leading order dependence
of the velocity field on the radial variable.
This will be done in the next section, together with a comparison
of the linear growth rates between the model and the full 
Navier-Stokes calculation.
In the third section we analyze the nonlinear behavior. First we 
investigate possible scaling solutions of the model equations, then
we compare with numerical simulations. In the final section, 
we present some tentative conclusions.

\section{The model}
In our derivation of the slender jet model we closely follow
\cite{ED94}. The Navier-Stokes equation for an incompressible 
fluid of viscosity $\nu$ read in cylindrical coordinates:
\[
\partial_t v_r + v_r\partial_rv_r + v_z\partial_zv_r - v_{\phi}^2/r = 
-\partial_r p/\rho 
+ \nu(\partial^{2}_{r}v_r + \partial^{2}_{z}v_r + \partial_rv_r/r -
v_r/r^2), 
\]
\[
\partial_t v_z + v_r\partial_rv_z + v_z\partial_zv_z =
-\partial_z p/\rho 
 + \nu(\partial^{2}_{r}v_z + \partial^{2}_zv_z + \partial_rv_z/r), 
\]
\[
\partial_t v_{\phi} + v_r\partial_rv_{\phi} + v_z\partial_zv_{\phi}
+ v_r v_{\phi}/r = 
\nu(\partial^{2}_{r}v_{\phi} + \partial^{2}_zv_{\phi} + 
\partial_rv_{\phi}/r - v_{\phi}/r^2),
\]

with the incompressibility condition
\[
\partial_rv_r + \partial_z v_z + v_r/r = 0 .
\]
Here we have assumed that the velocity field does not depend on 
the angle $\phi$. Exploiting incompressibility, $v_z$ and 
$v_r$ can be expanded in a power series in $r$:

\begin{eqnarray}\label{exp}
&& v_z(z,r) = v_0(z) + v_2(z) r^2 + \dots \\
\nonumber
&& v_r(z,r) = -\frac{v_0'(z)}{2} r - \frac{v_2'(z)}{4} r^3 - \dots .
\nonumber
\end{eqnarray}
Here and in the following a prime refers to differentiation 
with respect to $z$. 

The crucial trick to make an expansion in $r$ work in the presence 
of rotation is to rewrite $v_{\phi}(z,r)$ in terms of the angular
momentum per unit length $\ell(z)$ of the corresponding solid 
body rotation:
\begin{equation}\label{phiexp}
v_{\phi}(z,r) = \frac{2\ell(z)}{\pi\rho h^4(z)} r + b r^3 + \dots .
\end{equation}
Here $h(z)$ is the local thread radius, hence no overturning of
the profile is allowed. Just as without rotation, the equation of
motion for $h(z,t)$ follows from mass conservation based on the 
leading order expression for $v_z$:
\begin{equation}\label{hequ}
\partial_t h + v_0 h' = - v'_0 h/2 \quad.
\end{equation}
Finally, the pressure is expanded according
to 
\begin{equation}\label{pexp}
p(z,r) = p_0(z) + p_2(z) r^2 + \dots .
\end{equation}
Plugging this into the equation of motion for $v_r$, to leading order in 
$r$ one finds the balance 
\begin{equation}\label{p2}
p_2 = \frac{2 \ell^2}{\pi^2 \rho h^8},
\end{equation}
while the leading balance for the $v_z$-equation remains 
\begin{equation}\label{vzexp}
\partial_t v_0 + v_0 v'_0 = -p'_0/\rho + \nu(4v_2 + v''_0). 
\end{equation}
Lastly, the $v_{\phi}$-equation leads to 
\begin{equation}\label{lexp}
\partial_t \ell + \ell v'_0 + 4\ell v_0h'/h + v_0h^4(\ell/h^4)' = 
\nu h^4(4\pi \rho b + (\ell/h^4)'')
\end{equation}
to leading order. 

Equations (\ref{p2})-(\ref{lexp}) contain the unknown functions $p_0$,
$v_2$, and $b$ which need to be determined from the boundary 
conditions. The normal stress balance ${\bf n \sigma n} = \gamma
\kappa$ gives 
\[
p_0 + p_2 h^2 = \gamma \kappa - v'_0,
\]
where $\kappa$ is the sum of the principal curvatures. 
As in the case without rotation, the tangential stress balance
${\bf n \sigma t} = 0$ gives
\[
-3v'_0 h' - v''_0 h/2 + 2v_2 h = 0
\]
for ${\bf t}$ pointing in the axial direction, but a new condition 
\[
\pi\rho hb = h'(\ell/h^4)'
\]
for ${\bf t}$ pointing in the azimuthal direction. Putting this 
together, one is left with a closed equation for $h$, $v_0$, 
and $\ell$:
\begin{eqnarray}\label{me}
&& \partial_t h + v h' = -v' h/2 \\
\nonumber
&& \partial_t v + v v' = -\frac{\gamma}{\rho}\kappa' + 
\frac{2}{\rho^2\pi^2} (\ell^2/h^6)' + 3\nu(v'h^2)'/h^2 \\
&& \partial_t l + (v l)' = \nu (h^4(\ell/h^4)')' ,
\nonumber
\end{eqnarray}
where we have dropped the subscript on $v_0$. The same equations
were derived independently by Horvath and Huber \cite{H99}.

The most obvious way to test this model is to compare with
the known results for the stability of the full Navier-Stokes
equation. To that end we linearize (\ref{me}) about a solution 
with constant radius $r_0$ and rotation rate $\Omega$:
\begin{eqnarray}\label{an}
&& h(z,t) = r_0(1 + \epsilon e^{\omega t} \cos(kz)) \\
\nonumber
&& v(z,t) = -2\epsilon\frac{\omega}{k} e^{\omega t} \sin(kz) \\
&& \ell(z,t) = \frac{\pi}{2}\rho\Omega r_0^4
(1 + \epsilon e^{\omega t}\alpha \cos(kz)) .
\nonumber
\end{eqnarray}
Eliminating $\alpha$, this leads to the equation
\begin{equation}
\bar{\omega}^3 + \frac{4\bar{k}^2}{Re} \bar{\omega}^2 + \frac{\bar{k}^2}{2}
(-1+\bar{k}^2+L^{-1}+6\bar{k}^2/Re^2) \bar{\omega} + 
\frac{\bar{k}^4}{2 Re}(-1+\bar{k}^2-L^{-1}) = 0,
\label{poly}
\end{equation}
where $\bar{k} = kr_0$ and 
$\bar{\omega} = \omega(\rho r_0^3)^{1/2}/\gamma^{1/2}$ are
dimensionless.
We have introduced the Reynolds number 
$Re = (\gamma r_0)^{1/2}/(\rho^{1/2}\nu)$, based on a balance of 
capillary and viscous forces. Note that this convention differs 
from that of \cite{RJ70}. Putting $\bar{\omega}=0$ one reproduces
the exact stability boundaries (\ref{stab}). However 
one can see that the inviscid limit $Re\rightarrow \infty$ is
a very singular one, in disagreement with the full solution.
Namely, for this limit one finds the three branches
\begin{equation}
\bar{\omega}_{1/2}^2 = \frac{\bar{k}^2}{2}(1-\bar{k}^2-L^{-1}), \quad 
\bar{\omega}_3 = Re^{-1} 
\frac{\bar{k}^2(1-\bar{k}^2+L^{-1})}{\bar{k}^2-1+L^{-1}}.
\label{branch}
\end{equation}
Thus $\bar{\omega}_3$ is the only unstable mode in the range 
$1-L^{-1} < \bar{k}^2 < 1+L^{-1}$, but goes to zero when the viscosity 
becomes small. The reason for this behavior, which is not 
found in the solution of the full equations, lies in the 
appearance of a very thin boundary layer for small
viscosities \cite{P67}. Namely, Rayleigh's stability 
criterion for a rotating fluid implies that the interior of the 
fluid is {\it stabilized}. This forces any disturbance to be confined
to a boundary layer of thickness 
\[
\delta = \frac{\omega}{2\Omega k}
\]
near the surface of the jet, and $\delta$ becomes very small
for $\bar{k} \approx 1$. But this additional length scale 
is not captured by our slender jet expansion. Only for 
high viscosities is the boundary layer smoothed
out sufficiently, and from (\ref{poly}) one finds the dispersion relation
\begin{equation}
\bar{\omega} = \frac{Re}{6}(1-\bar{k}^2+L^{-1}),
\label{high}
\end{equation}
which is consistent with the full theory in the limit of
small $\bar{k}$. 

\section{Nonlinear behavior}
Our main interest lies of course in the behavior close
to pinch-off. Close to the singularity, one expects 
the motion to become independent of initial conditions, 
so it is natural to write the equations of motion 
in units of the material parameters of the fluid alone. 
In addition to the known \cite{PSS} units of length and time,
$\ell_{\nu} = \nu^2\rho/\gamma$ and $t_{\nu} = \nu^3\rho^2/\gamma^2$,
one finds an angular momentum scale $\ell_0 = \nu^5\rho^2/\gamma^2$.
Note that this scale is only about $1.9\cdot10^{-14}$ g cm/s for
water, corresponding to a frequency of $2\cdot 10^{-11} s^{-1}$
for a 1 mm jet, 
so even the smallest amount of rotation will be potentially 
relevant. Rewriting the equations of motion (\ref{me}) in 
units of $\ell_{\nu}, t_{\nu}$, and $\ell_0$, one can effectively 
put $\rho=\nu=\gamma=1$, leading to a universal form of the
equations, independent of the type of fluid. 

In addition, one can look for similarity solutions \cite{E}
of the form
\begin{eqnarray}\label{sim}
&& h(z,t) = t'^{\alpha_1}\phi(z'/t'^{\beta}) \\
\nonumber
&& v(z,t) = t'^{\alpha_2}\psi(z'/t'^{\beta}) \\
&& \ell(z,t) = t'^{\alpha_3}\chi(z'/t'^{\beta}) ,
\nonumber
\end{eqnarray}
where $t' = t_0 - t$ and $z' = z-z_0$ are the temporal and
spatial distances from the singularity where $h$ goes to zero.
We have assumed that everything has been written in units of the
natural scales $\ell_{\nu}, t_{\nu}$, and $\ell_0$.
By plugging (\ref{sim}) into the equations of motion, 
and looking for solutions that balance the $t'$-dependence,
one finds a unique set of values for the exponents:
\[
\alpha_1 = 1,\quad \alpha_2 = -1/2,\quad \alpha_3 = 5/2,\quad \beta=1/2.
\]
In addition, one obtains a set of three ordinary differential 
equations for the similarity functions $\phi, \psi$, and $\chi$.
So far we have not been able to find consistent solutions to
these equations, which match onto a solution which is static 
on the time scale $t'$ of the singular motion. This is a necessary 
requirement since the fluid will not be able to 
follow the singular motion as one moves away from the pinch point.

This negative result is consistent with simulations of the 
equations for a variety of initial conditions. To avoid spurious 
boundary effects, we considered a solution of (\ref{me}) with
periodic boundary conditions in the interval $[-1,1]$ and an
additional symmetry around the origin. This ensures that the 
total angular momentum is conserved exactly. We took the fluid
to be initially at rest and the surface to be 
\begin{equation}
h_{init}(z) = r_0(1 + 0.3 \cos(2\pi x)),
\label{init}
\end{equation}
with $r_0 = 0.1$. The angular momentum was distributed 
uniformly with the initial value $\ell_{init}$, corresponding
to 
\[
L = \frac{\pi^2}{4}\frac{\gamma\rho r_0^5}{\ell^2_{init}} .
\]

\begin{figure}
  \begin{center}
    \leavevmode
    \epsfsize=0.4 \textwidth
    \epsffile{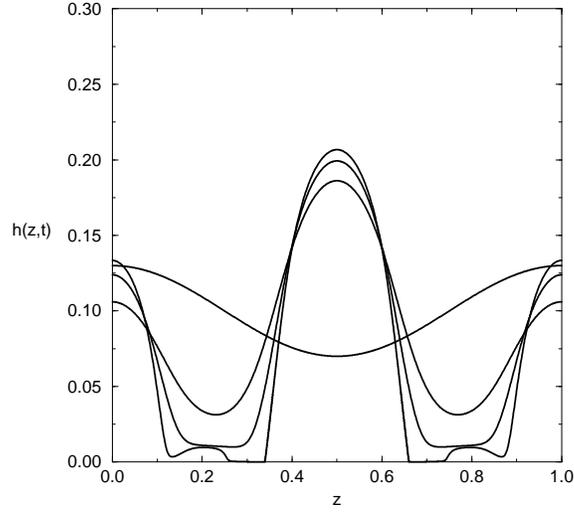}
    \caption{
The height profile in a numerical simulation with
$Re = 4.5$ and $L = 0.25$. Shown are the initial 
condition, and the times when the minimum has reached 
$h_{min} = 10^{-1.5}, 10^{-2}$, and $10^{-5}$, 
at the end of the computation.
                }
    \label{fig1}
  \end{center}
\end{figure}

\begin{figure}
  \begin{center}
    \leavevmode
    \epsfsize=0.6 \textwidth
    \epsffile{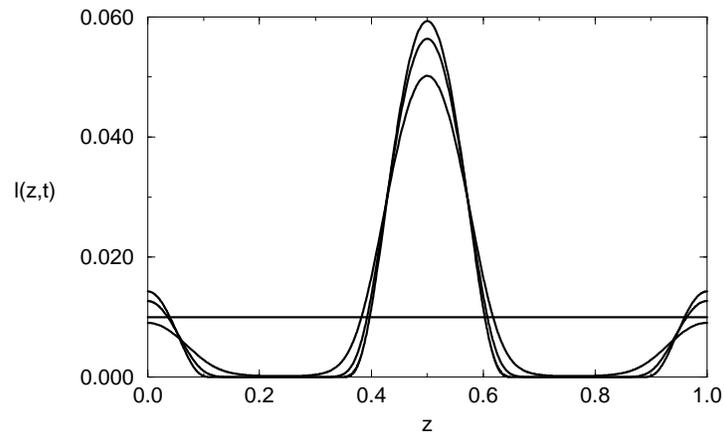}
    \caption{
The angular momentum profiles $\ell(z,t)$
corresponding to Fig. \protect{\ref{fig1}}.
                }
    \label{fig2}
  \end{center}
\end{figure}

\begin{figure}
  \begin{center}
    \leavevmode
    \epsfsize=0.6 \textwidth
    \epsffile{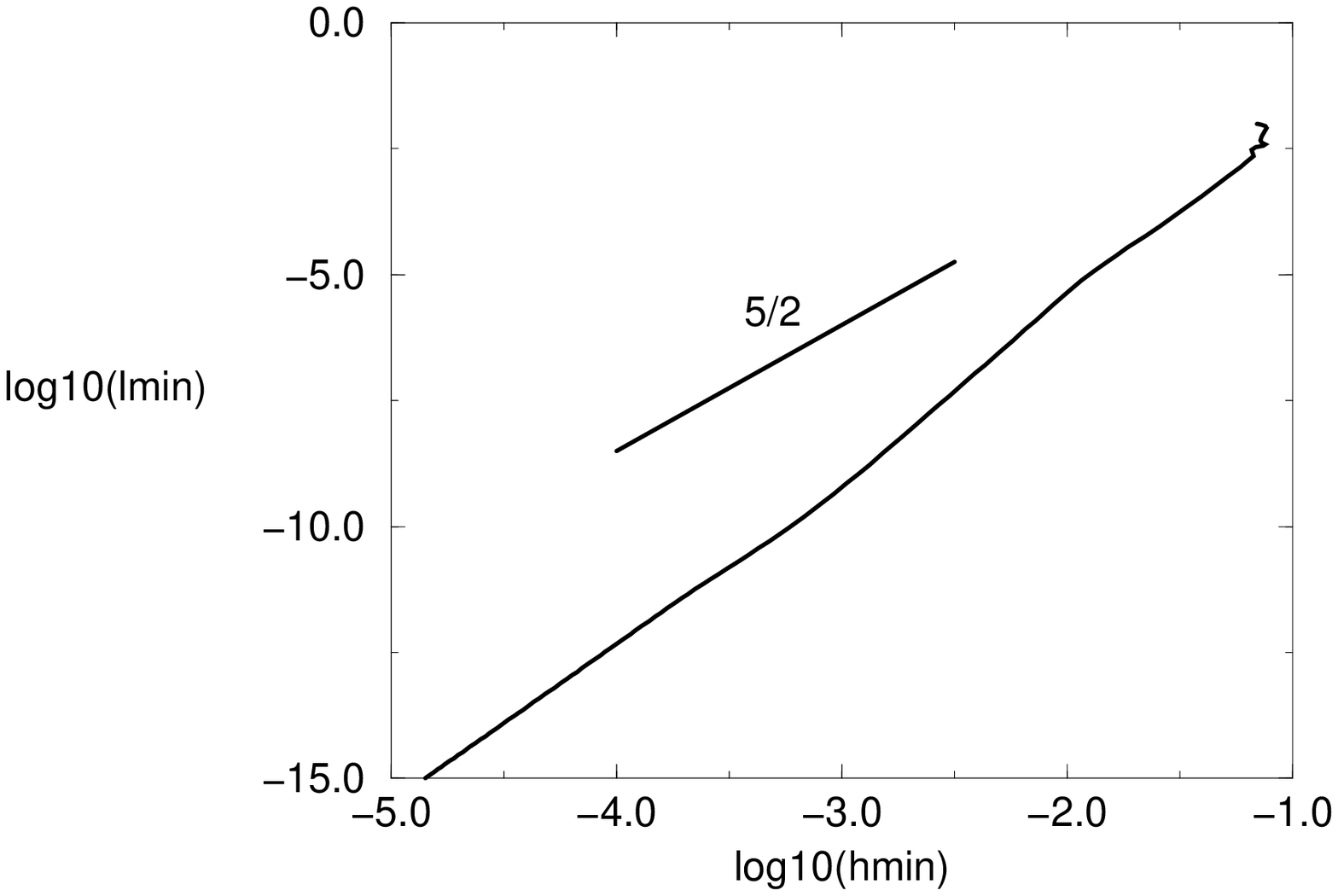}
    \caption{
The minimum value of the angular momentum as function of the 
minimum height. It is found that $\ell_{min}$ decreases faster than
$h_{min}^{5/2}$, which would exactly balance surface tension 
and centripetal forces. 
                }
    \label{fig3}
  \end{center}
\end{figure}

Figures \ref{fig1} and \ref{fig2} show a numerical simulation of (\ref{me})
with $Re = 4.5$ and $L = 0.25$ using a numerical code very 
similar to the one described in \cite{BEJNS}. Written in units of 
the intrinsic angular momentum scale, $\ell_{init}/\ell_0 = 6\cdot 10^3$,
so $\ell$ is potentially relevant. The thread pinches
on either side of the minimum, pushing fluid into the center. 
As seen in the profiles of $\ell$, the angular momentum is expelled very 
rapidly from the regions where $h$ goes to zero and also 
concentrates in the center. This is 
confirmed by a plot of the minimum of $\ell$ versus the 
minimum of $h$. On the basis of the similarity solution 
(\ref{sim}), a power law $\ell_{min} \propto h_{min}^{5/2}$
is to be expected. Instead, Fig. \ref{fig3} shows that 
$\ell_{min}$ decays more rapidly, making angular momentum 
asymptotically irrelevant. Indeed, a comparison of the 
similarity function $\phi$ as found from the present simulation 
shows perfect agreement with the scaling theory in the 
absence of rotation \cite{E}. The behavior of $\ell_{min}$ 
should in principal be derivable from the linear equation 
for $\ell$ with known time-dependent functions $h(z,t)$ and $v(z,t)$. 
Unfortunately, $\ell_{min}$ does not seem to follow a simple
scaling law except below $h = 3\cdot 10^{-4}$, where the power is
close to 3.13. It is not clear how to extract this power analytically.

\begin{figure}
  \begin{center}
    \leavevmode
    \epsfsize=0.6 \textwidth
    \epsffile{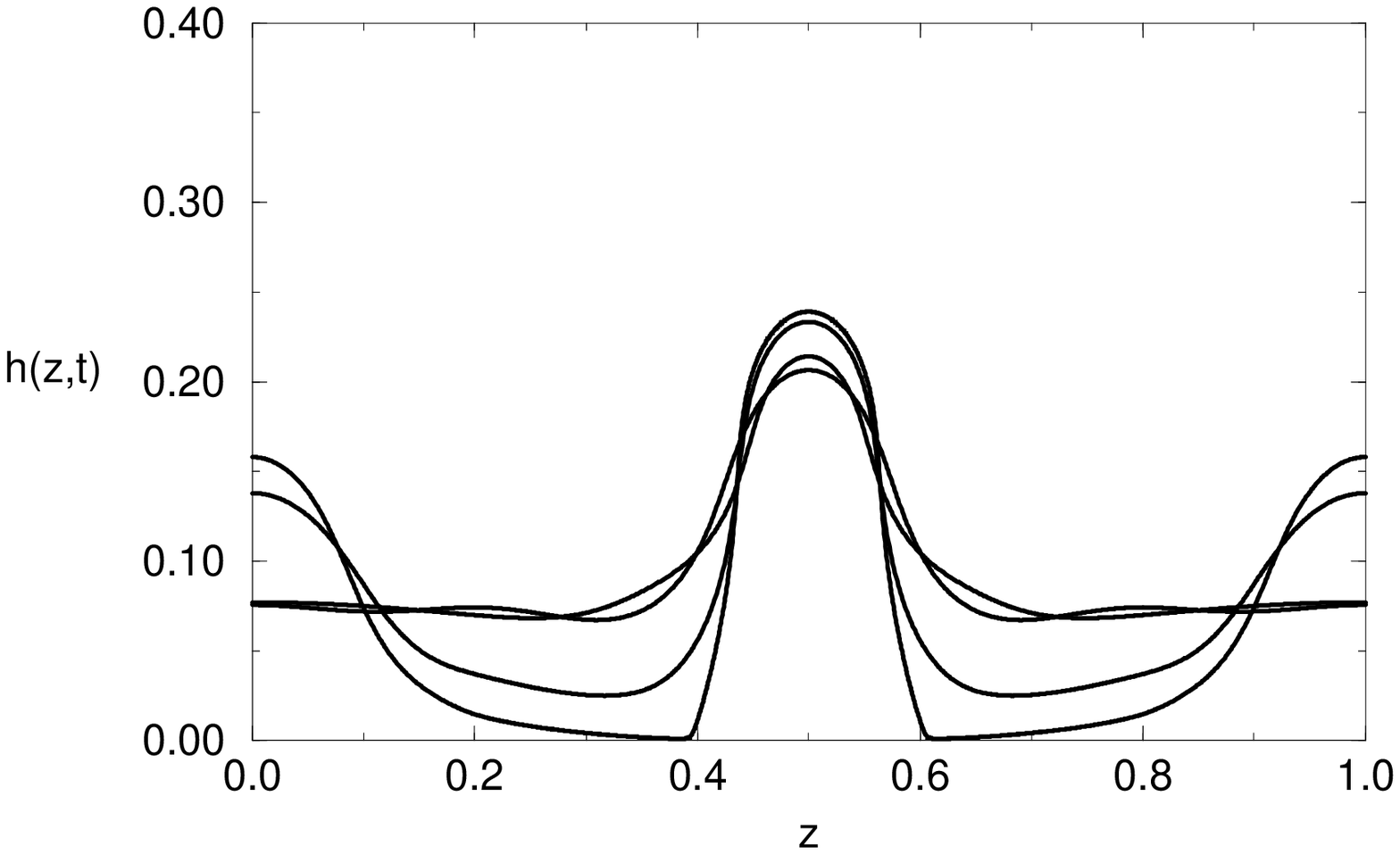}
    \caption{
A numerical simulation with twice 
the angular momentum of Fig. \protect{\ref{fig1}}. 
The height profiles are shown in time intervals of 0.05 
and at the end of the simulation. Centripetal forces draw
the fluid out into a disc.
                }
    \label{fig4}
  \end{center}
\end{figure}

One might think that by increasing the angular momentum the 
system would cross over to a different behavior. To test this, 
the initial angular momentum was doubled to give $L=0.0625$.
At $L = 0.5$ centripetal and surface tension forces are 
balanced, so decreasing $L$ significantly below this value
will cause rotation to be important initially.
Indeed, instead of pinching down immediately, the fluid is first
pulled into a narrow disc, while the radius of the surrounding
fluid remains constant, cf. Fig. \ref{fig4}. 
Eventually
this outward motion stops, as surface tension and
centripetal forces reach an equilibrium. Only then does the fluid 
pinch down at the edge of the disc. The behavior close to the 
pinch point is however exactly the same as for smaller angular
momentum. As a word of caution, one must add that our model
is certainly not valid at the edges of the disk, where slopes
become very large. In fact, the very sharp gradients encountered 
in this region may be due to the fact that the fluid really 
wants to develop {\it plumes}. As is observed for low viscosity
\cite{BEJNS}, the viscous terms prevent the interface from 
overturning, but at the cost of producing unrealistically sharp 
gradients. 

\section {Conclusions}
The present investigation is only a first step towards the
understanding of the role of rotation in droplet pinching. 
A major challenge lies in finding a description valid at 
low viscosities. This can perhaps be done by incorporating
the boundary layer structure near the surface into the slender
jet approximation. The relevance of this lies in the fact 
that angular momentum is potentially more important at low
viscosities, when there is less frictional transport out 
of the pinch region. In fact it can be shown that the inviscid 
version of (\ref{me}) does not describe breakup at all, since 
centripetal forces will always dominate asymptotically. This 
result is of course only of limited use since the model equations 
are definitely flawed in that regime. 

In addition, there remains the possibility that 
a region in parameter space exists where angular momentum 
modifies breakup even at finite viscosities. We cannot 
make a definite statement since the additional 
variable makes it hard to scan parameter space completely.
Finally, spinning jets have not received much attention in
terms of experiments probing the non-linear regime. The discs
found at high spinning rates (cf. Fig. \ref{fig4}) 
are a tantalizing new feature, and
to our knowledge have not been found experimentally.
The lowest value of $L$ reported in \cite{RJ70} is 0.43, 
which is even larger than the value of Fig. \ref{fig1}.
However, 0.0625 would easily be reachable by increasing
the jet radius.

\acknowledgements
The authors are indebted to Sid Nagel for pointing out this
problem and for stimulating discussions. J.E. thanks Howard
Stone for his hospitality, which he has shown in so many 
ways, and for stimulating discussions. J.E. is also grateful
to Leo Kadanoff and the Department of Mathematics at the 
University of Chicago, where this paper was written, for
making this another enjoyable summer. M.B. acknowledges 
support from the NSF Division of Mathematical Sciences,
and the A.P. Sloan foundation. J.E. was supported by the 
Deutsche Forschungsgemeinschaft through SFB237.

\end{document}